\documentclass[preprint,12pt]{elsarticle}

\usepackage{amssymb}
\usepackage{graphicx}
\usepackage[title]{appendix}
\usepackage{booktabs}
\usepackage{array}
\usepackage{longtable}
\usepackage{comment}
\usepackage{times}
\usepackage{makecell}
\usepackage{multirow}
\usepackage{textcomp}
\usepackage{rotating}
\usepackage{lscape}
\usepackage{xcolor}
\usepackage{lscape}

\AtBeginDocument{%
  \providecommand\BibTeX{{%
    \normalfont B\kern-0.5em{\scshape i\kern-0.25em b}\kern-0.8em\TeX}}}

\journal{Information and Software Technology}

\begin{document}

\begin{frontmatter}

\title{Human Error Management in Requirements Engineering: Should We Fix the People, the Processes, or the Environment?}

\author[inst1]{Sweta Mahaju}

\affiliation[inst1]{organization={Department of Computer Science},%Department and Organization
            addressline={Middle Tennessee State University}, 
            city={Murfreesboro}, 
            state={TN},
            country={USA}}

\author[inst2]{Jeffrey C. Carver}

\affiliation[inst2]{organization={Department of Computer Science},%Department and Organization
            addressline={The University of Alabama}, 
            city={Tuscaloosa}, 
            state={AL},
            country={USA}}
            
\author[inst3]{Gary L. Bradshaw}
            
\affiliation[inst3]{organization={Department of Psychology},%Department and Organization
            addressline={Mississippi State University}, 
            city={Starkville}, 
            state={MS},
            country={USA}}

\begin{abstract}
\textit{Context:} Software development is a human-centric activity and hence vulnerable to human error. 
Human errors are errors in the human thought process.
To ensure software quality, it is important for practitioners to understand how to manage these human errors. 
Organizations often introduce changes into the requirements engineering process to either \textit{prevent} human errors from occurring or to \textit{mitigate} the harm caused when those errors do occur.  
While there are studies on human error management in other disciplines, research on the prevention and mitigation of human errors in software engineering, and requirements engineering specifically, are limited.
The current studies in software engineering do not provide strong results about the types of changes that are most effective in requirements engineering.
\textit{Objective:} The goal of this paper is to develop a taxonomy of human error prevention and mitigation strategies based on data gathered from requirements engineering professionals. 
\textit{Method:} We performed a qualitative analysis of data from two practitioner surveys on requirements engineering practices to identify and classify strategies for the prevention and mitigation of human errors.
\textit{Results:} We organized the human error management strategies into a taxonomy based on whether the changes primarily affect People, Processes, or the Environment.
Inside each of these high-level categories, we further organized the strategies into low-level classes.
The results show more than 50\% of the reported strategies require a change in Process, 23\% require a change in Environment, 21\% require a change in People, with the remaining 5\% too ambiguous to classify.
In addition, more than 50\% of the strategies focus on Management activities of requirements engineering.
\textit{Conclusions:}
The Human Error Management Taxonomy provides a systematic classification and organization of strategies for prevention and mitigation of human errors in requirements engineering.
This systematic organization provides a foundation upon which research can build.
\end{abstract}


\end{frontmatter}

\section{Introduction}
\label{sec:Introduction}
Requirements engineering is the foundational phase of the software development life-cycle (SDLC). 
During this phase, people interact with one another to elicit, document, and validate software requirements. 
The involvement of people in this phase provides ample opportunity for human error. 
\textit{Human errors} are failings in the human thought processes that, if unaddressed, can lead to faults in requirements documents. 
Human errors occur when requirements engineers misunderstand customer needs, apply the wrong type of solution, or make a mistake due to memory overload~\cite{reason1990human}. 
These types of human errors are particularly problematic because their effects can remain hidden for a long time and often only surface as failures much later in the software development process~\cite{viller1997human}. 
As a result, frequent rework and changes in requirements can lead to more errors, degrade software quality, and increase costs~\cite{askarinejadamiri2017impact}. 

When a human error, or any other type of sub-optimal process, occurs, a team must either take action to mitigate the potential effects of the error or introduce changes to prevent similar errors in the future.
Even though this topic is important for improving software quality, we could find no published systematic studies of these types of prevention and mitigation techniques in requirements engineering.  
To address this deficiency in the literature, we need to answer questions like: \textit{How can we organize prevention and mitigation approaches 
into a framework that reveals their underlying structure?} and \textit{Can we identify particular types of prevention and mitigation approaches that are likely to be effective or ineffective?} 
Answers to these types of questions will help software development organizations proactively implement appropriate error prevention and mitigation techniques as well as avoid the painful process of learning from their own mistakes.  
Review of the broader literature also suggests that certain types of error mitigation and prevention strategies are unlikely to be effective, while other techniques are more likely to prove efficacious.

Human error management improves software quality through use of error prevention and error mitigation activities. 
Error prevention activities reduce the likelihood of errors through changes to the development process.  
Error mitigation activities seek to minimize the downstream effects of errors after they occur.
However, given the lack of studies on error management approaches, human error still persists in requirements engineering.
Nevertheless, individual developers have made efforts to prevent errors, even without benefit of published research that analyzes effective error prevention and mitigation strategies. 
Therefore, researchers and practitioners can benefit from a formal analysis of the effectiveness of error prevention and mitigation strategies.
In addition, a classification of error prevention and mitigation strategies will provide a foundation upon which future research can build. 
Therefore, the primary research objective of this paper is \textit{\textbf{to construct a human error management taxonomy and to classify error prevention and mitigation strategies across the stages of the requirements engineering process}}.

To address this goal, we analyzed data from surveys of professional requirement engineers~\cite{hu2018using,fernandez2017naming}. 
These surveys gathered information about problems during the requirement engineering process along with the approaches to manage those problems~\cite{fernandez2017naming}. 
In our previous work, two of the authors analyzed this data to identify specific error prevention and mitigation approaches practitioners employed to address different types of human errors~\cite{hu2018using}. 
The research was focused on extracting specified prevention or mitigation strategies for each type of error and labeling each strategy as either \textit{prevention} \textit{mitigation} or \textit{both}.

In the current study, we performed a much deeper analysis of the prevention and mitigation strategies extracted in the previous work.
The output of the previous is the ``raw data'' used as input to the work described here.
While the previous work provided a list of strategies and whether they were primarily for prevention or for mitigation, that work did not analyze those strategies at a deeper level to systematize them.
This paper follows up on that work and systematically builds a taxonomy of prevention and mitigation strategies along with the requirements activity in which those strategies are most useful.
To implement the approach, we followed a process consisting  of individual analysis and coding followed by discussion, as suggested by Kitchenham et al.~\cite{kitchenham2010can}. 
As a result, we attain the goals of developing an organized structure of error management strategies in this study.

The primary contributions of this paper are:
\begin{enumerate}
    \item An analysis of human error prevention and mitigation strategies used in industry, 
    \item An analysis of which of these activities occur in each requirements engineering activity, and
    \item A classification of those strategies into a taxonomy that includes their requirements engineering activities.
\end{enumerate}

\section{Research Background}
\label{sec:Background}
This section provides background research on human error and on error prevention and mitigation in software engineering and other domains.

\subsection{Human Error Research}
A \textit{Human Error} is an error in the human thought process which results in deviations from expected actions producing unexpected or undesired outcomes~\cite{reason1990human}. 
People experience human error in their daily lives. 
Some simple examples of human error include: 
(1) going into a room and then forgetting the reason why, 
(2) throwing eggs into the trash instead of the eggshells, and 
(3) searching for your glasses when they are on your head. 
In requirements engineering, as mentioned in Section~\ref{sec:Introduction}, human error occurs due to misunderstanding or miscommunication between the requirements engineers and the customers. 
Examples of human errors in requirements engineering are typos, missing requirements, missing information, and wrong assumptions~\cite{anu2018development}.

Prior research on human error in other domains has shown that attempting to resolve human error by encouraging changes in the people involved is not an effective~\cite{gosbee2003human,kellogg2017our,wu2008effectiveness} approach. 
For example, Kellogg et al. investigated and analyzed the sources of human error in the health-care domain over eight years~\cite{kellogg2017our}. 
They identified and studied the effectiveness of the most frequent solutions to handle human errors. 
They concluded that solutions that focus on changing people, such as reminding them not to make mistakes, are seldom effective because the circumstances that promote errors remain intact. 
We anticipate the same result will occur as companies and organizations implement prevention and mitigation practices in requirements engineering. 
For example, in the case of personnel turnover, when a requirements engineer who was trained to prevent an error leaves the company, the replacements could replicate the same error.

Kellogg et al. further stressed that the error management approaches that change the technologies, tools, or organizational policies are more useful, yet are not commonly proposed solutions~\cite{kellogg2017our}. 
The disastrous ``landing gear incidents'' from World War II provide an example of the claim. 
More than 2000 aircraft accidents occurred during the first two years of World War II because the pilots in command mistook the flap controls for landing gear controls while landing. 
Planes landed with their gear still up, often damaging the plane extensively.
Subsequent investigation found that human error occurred because both the landing gear and the flap lever were precisely the same (in shape, size, method of operation) and located together. 
After redesigning the gear and the lever as a mitigation measure, it helped to minimize the accidents. 
The example showed that even correctly trained pilots are vulnerable to human error in harsh conditions~\cite{hendrick2002ergonomic}. 
The human error experts realized that it is better to focus on designing better machines rather than training people to operate poorly designed machines and hoping for the best outcome. 
Similarly, many of the changes implemented as problem-solving strategies in requirements engineering use ineffective measures that may not help them to solve the problem.

\subsection{Human Error Research in Requirements Engineering}
To help researchers and practitioners better understand the types of human errors that occur during requirements engineering, Anu, et al., conducted a systematic literature review to identify, organize, and structure the human errors described in the literature~\cite{anu2018development}.
Based upon a standard taxonomy of human errors from Cognitive Psychology~\cite{reason1990human}, they created the Human Error Taxonomy (HET) to organize the human errors in requirements engineering. 
The HET extends Reason's human error classes of Slips, Lapses and Mistakes and provides a framework which describes the types of human errors requirements engineers can make.
An understanding of the various type of human errors can provide insight into the most appropriate type of prevention and mitigation approaches.

\subsection{Error Prevention and Mitigation Approaches in Software Engineering}
\label{sec:EPM}
Huang, et al. developed a human error taxonomy by investigating human errors committed by software engineers in the China Aviation Software Industry~\cite{huang2012taxonomy}.
They developed this taxonomy to help prevent human error in software engineering.
As a result of the study, the authors proposed an error prevention framework DPeHE (Defect Prevention approach based on Human Error mechanisms)~\cite{huang2017software}.  
DPeHE is comprised of two modules to enhance the practitioner's meta-cognitive capabilities - knowledge training and regulation training.  
The knowledge training module can help practitioners familiarize themselves with \textit{why}, \textit{how}, and \textit{what} kind of errors they commit. 
The regulation training module can help practitioners practice problem-solving activities using a root cause identification approach based on the knowledge gained through the knowledge training. 
A series of case studies illustrated the effectiveness and usefulness of the DPeHE framework in preventing errors by enhancing software developers' meta-cognitive ability.
However, because this work used only two companies to evaluate the framework, there are questions about the generalizability of its application and effectiveness.

Huang also studied human errors in post-completion activities~\cite{huang2016post}.
Post-completion errors are those that occur when a practitioners does not complete a sub-task after completion of a primary task. 
An example of a post-completion error is forgetting to retrieve your card from an ATM (Automated Teller Machine) after completing your transaction~\cite{huang2016post,back2007does}.
Huang conducted an experiment in a programming contest that contains a sub-task to be completed at the end of the primary task.
Out of 55 students who participated in the study, 23 omitted the sub-task.
She found that students are vulnerable to committing human error despite being motivated and trained to avoid them. 
She also emphasized people-related solutions to human error are ineffective. 
Huang proposed three strategies to defend against post-completion errors (1) eliminate the unnecessary sub-task, (2) change the procedure of the task, or (3) highlight the task in bright color or bold font.
These strategies all focus on changes to processes or procedures.
While the results of this study are important, the participants were students rather than practitioners, potentially reducing the external validity.

Firesmith identified the most common requirements engineering problems, explained their impacts, and suggested a possible solution~\cite{firesmith2007common}. 
The author described the requirements problems and their causes thereby illustrating human errors behind those problems. 
The main goal of the paper was to provide guidelines for handling requirements problems. 
While readers with knowledge of human error will likely understand the implicit explanation of the proposed human error management approaches, which could help them be prepared to address the human errors, the paper itself does not specifically describe human error. 
Therefore, the limitation of the paper, relative to human error, is that it depends on the practitioners' ability to identify human errors from the requirements problems explained in the paper. 
Moreover, the study lacked a strong cognitive perspective and was based on the authors' experience and opinion.

Lopes et al. analyzed the importance of error types in the context of requirements problems, with the help of requirements engineers~\cite{lopes2013application}. 
They developed an expert system to provide solutions for managing requirements errors.
The proposed system inputs a list of requirements problems and provides a list of the human error types, ordered by importance, that could have occurred and a suggestion of a common solution from the literature.
Because the expert system makes practitioners aware of the errors that cause the requirements problems, it acts as both an error prevention tool and an error mitigation tool.
However, the main limitation of this system is that the solutions provided come from the literature rather than from practice, raising questions about their validity in practice.

Walia and Carver conducted a systematic literature review to identify and classify human errors that occur during the requirements phase of software engineering~\cite{walia2009systematic}. 
Using this information they built the Requirements Error Taxonomy (RET).
Later, with the help of the error types in the RET, they conducted an empirical study using the RET to augment Lanubile et al.'s error abstraction approach~\cite{lanubile1998experimenting} for preventing errors in the requirements phase~\cite{walia2010evaluating}. 
Error abstraction is the retrospective process of tracing identified faults back to the underlying errors and using those identified errors to find more faults.
The result of the study showed that the participants who found more errors during requirements inspection made fewer errors during subsequent requirements creation, suggesting that the RET can be used as an error prevention tool~\cite{walia2010evaluating}.
While this study supported the validity of the RET as an error prevention tool, the RET itself is not based on a strong human error theory.

Hu et al. implemented the error abstraction process using error information from the Human Error Taxonomy (HET)~\cite{hu2018using,anu2018development}. 
They collected data related to requirements problems faced in industry, mapped them to the corresponding human errors, and classified the solutions the requirements engineers used to address the problems.
They categorized each solution as either a \textit{prevention} approach, i.e. preventing the error from occurring, or a \textit{mitigation} approach, i.e. managing the effects when the error does occur. 
They then mapped each solution to the corresponding error types. 
While this study can help practitioners be more aware of the prevention and mitigation strategies, the ability to recognize the appropriate strategy for each error type still depends upon the practitioner's ability.

This discussion illustrates that there is still a need for additional studies focused on the prevention and mitigation of human errors in requirements engineering, specifically, or software engineering, more generally. 
The existing studies have unaddressed gaps, including the lack of a strong underlying cognitive theory, generalizability and validity issues, dependency on developers' ability to identify the right error management strategies, and lack of a structured organization of error prevention and mitigation mechanisms.
Our current study seeks to address these gaps.

\section{Research Method}
\label{sec:Methodology}
This section describes the motivation for the study, the research questions, the data collection procedures, and an overview of the analysis method.
Because the analysis is iterative with each step building upon the results of the previous one, we describe the details of the analysis along with the results in Section~\ref{sec:Results}.

\subsection{Motivation for New Study}
The previous work (Section~\ref{sec:EPM}) described an initial analysis of data about prevention and mitigation of human errors.
This initial work still left a number of unaddressed gaps in need of further study.
Specifically, there is a lack of guidance to practitioners on how to identify which prevention or mitigation strategy is most appropriate for a given requirements problem.

In this paper, we use the same industrial datasets as the previous studies (described in Sections~\ref{sec:NAPIRE} and~\ref{sec:CAPS}) to conduct deeper analysis to organize the prevention and mitigation strategies identified in those studies into a taxonomy that contains detailed categories and sub-categories.
We also examine the specific requirements engineering activities within which each strategy is useful to provide additional information to help practitioners.

\subsection{Research Questions}
\label{sec:RQ}
The following research questions guide this study:
    \begin{description}
        \item[RQ1] What types of prevention and mitigation strategies do practitioners employ to address human errors in requirements engineering and how can we organize them into a formal taxonomy?
        \item[RQ2] Can we categorize the human error prevention and mitigation strategies into the taxonomy developed in RQ1?
    \end{description}
    
\subsection{Research Data}
\label{sec:Data}
To conduct our analysis, we used data from two existing datasets that contain requirements engineering information from industrial developers~\cite{hu2018using}.
These datasets include information on human error management strategies used by practitioners.
We combined these datasets to focus on the problems and solutions requirements engineers experienced in their projects.
The following sub-sections explain each dataset in more detail.

\subsubsection{Dataset 1: NaPiRE}
\label{sec:NAPIRE}
NaPiRE (Naming the Pain in Requirement Engineering) is a global initiative to study and understand the status quo of requirements engineering problems. 
As part of this initiative, an international group of requirements engineering researchers conduct a bi-annual survey on requirements engineering practices in industry. 
The results of this survey form a knowledge base of problems in requirements engineering~\cite{fernandez2017naming}. 
    
Here we use data from the 2016 survey because it was the last one where respondents described their strategies for handling requirements problems.
This survey has 226 respondents from 226 companies in 10 countries.
While the survey had a number of questions, for our work we focused on the answers to only one question.
The survey presented the participants with a list of 21 common requirements problems (identified in previous NaPiRE surveys) and asked them to select the 5 most common ones for their environment \cite{fernandez2017naming}.
~\ref{appendixA} provides a list of those requirements problems. 
For each of those problems, the respondents listed the causes, effects, and management strategies used to address them.
Because some of these requirements problems are related to human errors, these responses provide insight into prevention and mitigation strategies for human errors.

\subsubsection{Dataset 2: CAPS}
\label{sec:CAPS}
CAPS (The Center for Advanced Public Safety) is an interdisciplinary research organization at the University of Alabama that develops software for public safety applications.
CAPS provides a good setting to gather information about human errors in requirements engineering because its software development teams include requirements engineers, developers, and business analysts.
In previous work, two of the authors conducted a two-step survey of seven requirements engineers to gather information about the problems they face (some of which are related to human error) and the strategies they use to address those problems~\cite{hu2018using}.
Table~\ref{tab:tblCapsSurvey} shows the survey questions.

In the first survey, after explaining the concept of human errors, the researchers asked the questions from Survey 1 in Table~\ref{tab:tblCapsSurvey} to allow respondents to describe requirements problems and how they managed those problems.
Next, before the second survey, the authors introduced the Human Error Taxonomy (HET), its error types, and examples~\cite{anu2018development}.
We met with the participants face-to-face to introduce the HET. 
During this meeting we provided a detailed explanation of each high-level and low-level error class in the taxonomy~\cite{hu2018using}.
The researchers then asked the participants the questions from Survey 2 in Table~\ref{tab:tblCapsSurvey} to gather additional requirements problems the respondents could identify after learning about the HET.

\begin{table}[!htb]
    \caption{CAPS Survey Questions~\cite{hu2018using}}
    \label{tab:tblCapsSurvey}
% \usepackage{longtable}
% Note: It may be necessary to compile the document several times to get a multi-page table to line up properly

\begin{tabular}{c|p{.85\textwidth}}
\textbf{Survey} & \textbf{Questions} \\ \hline

\multirow{4}{*}{1} & What kinds of requirements problems have you encountered in your past software development processes? \\ \cline{2-2}
 & For those problems, what kinds of prevention strategies you have taken or plan to take to address these problems? \\ \hline
\multirow{4}{*}{2} & Based on the human error taxonomy that we introduced, are there additional human errors that you encountered which you did not report during the first survey? \\ \cline{2-2}
 & For the errors described in the  previous question, what kinds of prevention strategies you have taken or plan to take to address these problems? \\ \hline
\end{tabular}
\end{table}

\subsection{Overview of Analysis Method}
\label{sec:Analysis}

Our analysis consisted of three phases: 
(1) identification of types of prevention and mitigation strategies, 
(2) classification of prevention and mitigation strategies, and 
(3) discussion on disagreements.

We performed the first two phases sequentially.
The goal of the first phase was to analyze the dataset as a whole, identify patterns in the reported strategies, and finally extract error management categories.
Each of the researchers separately reviewed the strategies in the dataset to identify their own set of categories.
In this phase, we did not try to classify each strategy, but rather only to identify overall categories into which we could later classify individual strategies. 
To identify these categories, each of us independently identified commonalities among the approaches to identify our own set of higher-level groupings for the strategies.
After discussions to arrive at an agreement (see phase 3 below), we produced an agreed-upon set of high-level and lower-level categories.
Then the work proceed to the second phase where the main goal was to individually classify each strategy into the categories identified in phase 1.
Each researcher separately analyzed and classified each prevention and mitigation strategy into the error management categories and sub-categories identified in the first phase.

The third phase (Discussion on Disagreements) occurred multiple times throughout the first two phases.
After each classification activity, we had a discussion to resolve any disagreements.
These periodic discussions helped ensure we had a consistent and correct understanding of the classification approach.
At the end of each discussion, all authors reached agreement.

\begin{figure}[!htb]
  \centering
  \includegraphics[width=5.5in]{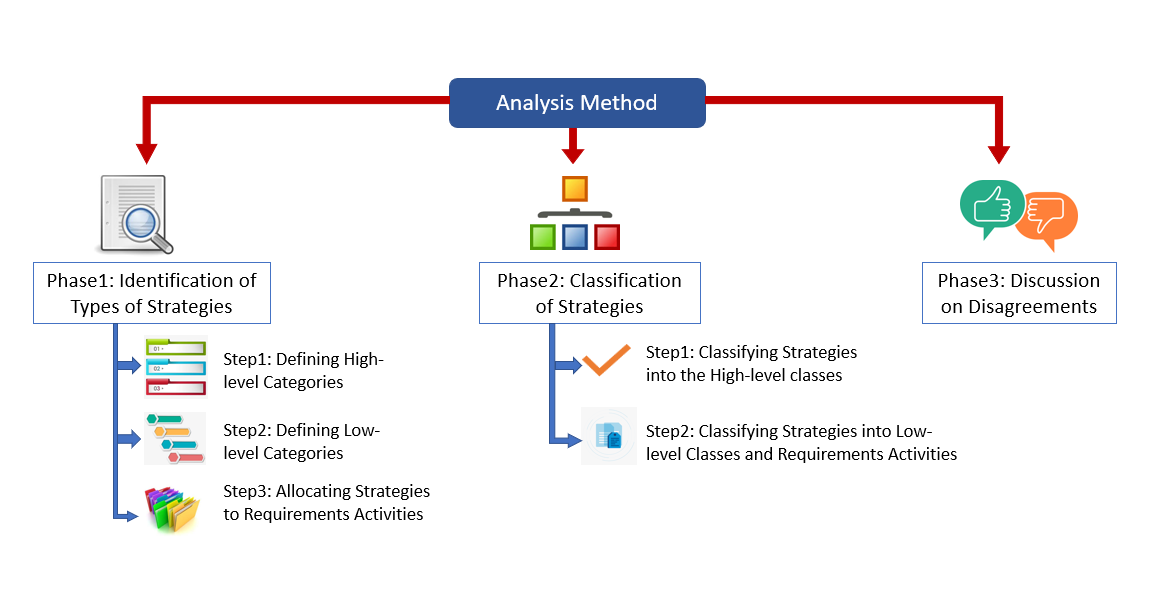}
  \caption{Classification of Prevention and Mitigation Strategies}
  \label{fig:AnalysisMethod}
\end{figure}

Figure~\ref{fig:AnalysisMethod} illustrates our method for classifying the prevention and mitigation strategies into high-level classes, low-level categories, and requirements activities.
Because the methodology for each step builds upon the results from the previous step, Section~\ref{sec:Results} combines the detailed description of the steps together with the results of each step.

\section{Results}
\label{sec:Results}
This section first provides an overview of the process we used to clean and prepare the data for analysis.
Then, the section describes the analysis process and results for each phase of the analysis.
We organize this section around the research questions posed in Section~\ref{sec:RQ}.

\subsection{Data Cleaning}
During the analysis process, we noticed that many of the reported prevention and mitigation strategies lacked the level of detail we would have liked.
To compensate for the lack of detail, we defined classification criteria. 
We observed some of the prevention and mitigation strategies contained two steps. 
For example, the strategy ``Introduction and use of a requirements quantification approach'' has two parts, ``introduction'' and ``use'', which are two different types of changes. 
According to our classification criteria, when strategies had two steps, like introduction and use, we split the strategy into two strategies.
As a result, we had 162 prevention and mitigation strategies to analyze. 

\subsection{RQ1: What types of prevention and mitigation strategies do practitioners employ to address human errors in requirements engineering and how can we organize them into a formal taxonomy?}
\label{sec:RQ1}

To answer this question, we conducted \textbf{Phase 1: Identification of Types of Prevention-Mitigation Strategies} of the analysis process.
The main goal of this phase was to identify categories into which we could classify the prevention and mitigation strategies.
Phase 1 has three Steps that build upon each other. 
We detail the steps and their results below.

\vspace{8pt}
\noindent
\textit{Step 1: Define high-level categories for prevention and mitigation strategies}

We began by attempting to identify a small set of high-level categories into which we could group the prevention and mitigation approaches.
Each author individually analyzed the data to identify an initial set of high-level categories for the observed prevention and mitigation strategies. 
Next, based on a discussion of these individual analyses, we agreed upon the high-level categories.
Then the first author reexamined the results of the individual analyses to organize them into the high-level categories and clarify the definitions of those categories.
Next, the other two authors reviewed the high-level categories' definitions and the initial results' organization into those categories and noted any disagreements.
After a second discussion, we agreed upon and finalized the following high-level categories: 

\begin{enumerate}
    \item \textbf{People}: 
    Changes introduced with the primary goal of modifying the behavior of the people. 
    Examples of these changes include training activities, counseling, and reminders. 
    Note that People changes directly impact specific individuals. 
    Therefore, when employee turnover occurs, the People changes are typically lost and have to be reintroduced for new employees.
\\    
    \item \textbf{Process}: 
    Changes made directly to the task flow or task procedures. 
    An example of a Process change is the introduction of a review stage that requires certain employees to sign-off on a document before it can proceed to the next phase. 
    Unlike the People changes, because they affect processes followed by all employees, Process changes should survive employee turnover.
    Process changes may occur either at the management level (i.e. greater oversight by management) or at the requirements engineering level (i.e. adding a a new step in the requirements engineering process).
\\
    \item \textbf{Environment}: 
    Changes to the environment within which the requirements engineering process occurs. 
    Examples of Environment change include introduction of new development tools or processes, addition of personnel with specialized expertise, or changes to the work culture.
    Environment changes often result in Process changes.
    For example, use of a new error-tracking system that changes the work environment also changes the development process. 
\\
    \item \textbf{Ambiguous}: 
    This category covers those changes that do not clearly fit into one of the above categories.
    Many of these Ambiguous changes resulted from descriptions that lacked detail necessary for us to clearly determine the type of the change and whether it belonged to one of the categories above.
\end{enumerate}

\vspace{8pt}
\noindent
\textit{Step 2: Define low-level categories for prevention and mitigation strategies}

After defining the high-level categories, the first author reviewed the prevention and mitigation strategies again to add additional low-level categories to describe specific types of changes that occur within each of the high-level categories.
The other two authors reviewed the low-level categories and their organization among the high-level categories.
They noted any disagreements or items in need of further discussion.
After discussion, the authors agreed on the low-level categories, as follows.

\begin{enumerate}
\item \textbf{People}

%Comment R4.12
\begin{enumerate}
    \item \textbf{Internal Training Activities (ITA)}:
    Changes in People to improve their skills and performance such as communication skills or time management skills. 
    This change occurs to the internal team members in a formal way. 
    For example, planning and execution of trainings (in order to improve skill and
    performance)
    
    \item \textbf{Internal Team Encouragement (ITE)}:
    Changes in People by encouraging the team as well as guiding them and making them aware of what needs to be done and how to do it right. 
    This is an informal guidance to the internal team members. 
    For example, Motivating team or empowering team; guiding team members to focus in customer requirements
    
    \item \textbf{External Training and Coaching Activities (ETA)}:
    Changes in People induced by increasing awareness or guidance to external stake holders. 
    This change occurs to the people external to the team, such as clients and customers in an informal way. 
    For example, guiding client to understand that developers do not necessarily have domain knowledge; Training/ Coaching/ Consulting: Customers need to be held coached on the importance of this stage of software development and how much it costs in support and maintenance to do otherwise.
\end{enumerate}

\item \textbf{Process}
\begin{enumerate}
    \item \textbf{Investigation (INV)}:
    Changes in Process by identifying what to do and how to do to solve the problem that occurred in the process. 
    For example- evaluation of tools or identification of process gaps. It is not an exact process change, but an investigation may cause a change in the process later. 
    It will not solve the problem by itself, but help try to identify what to do to solve the problem.
    
    \item \textbf{Refinement and Redesign (RR)}:
    Changes in Process induced by:a) changes in Environment due to the introduction of new processes, tools, or approaches; b) refining or modifying the existing process as needed; or c) implementing the existing process as a part of the process being modified. 
    For Example - Implementing a change management process, Prototypes, and wire-frames to help developers understand the requirements and use of the requirements quantification approach.
    
    \item \textbf{Definition of Roles and Responsibilities (DR)}:
    Changes in Process occurred by defining existing roles and responsibilities more clearly.
    For example, defining clear roles and responsibilities.
    
    \item \textbf{Resource Allocation (RA)}:
    Changes in Process by allocating correct resources in correct events. 
    For example, involving everyone representing every role in a software development project for the requirements gathering process of that project.
\end{enumerate}

\item \textbf{Environment}
\begin{enumerate}
    \item \textbf{Addition of New Process, Tool or Approaches (ANP)}:
    Changes in Environment by introducing new approaches, tools or process in software development.For example, introduction of agile methodology or new quality assurance process
    
    \item \textbf{Addition of Personnel (AP)}:
    Changes in Environment by addition of personnel that are not generally involved to work on that sub process. 
    Examples include requirement experts, end users, clients
    The sub-category AP is further divided into two categories:
    \begin{enumerate}
        \item[a]
        Internal (Int): Addition of personnel who work for the organization performing the requirement activities. For example, developers, managers, requirement engineers, business analysts and so on.
    \item[b]
        External (Ext): Addition of personnel who does not work for the organization performing requirements engineering activities. For example, end users, clients, external requirements experts.
    \end{enumerate}
    
    \item \textbf{Addition of Structure (AS)}:
    Change in Environment by defining or creating standards or templates to accommodate certain process. For example, defining quality criteria or creating specification templates.
    
    \item \textbf{Addition of New Role (AR)}:
    Changes in Environment by defining appropriate roles and responsibilities among existing team members. For example, appointing a team leader or introducing manager.
\end{enumerate}
\end{enumerate}

\vspace{8pt}
\noindent
\textit{Step 3: Allocate the low-level categories to phases of the requirements engineering lifecycle}

During the discussion in Step 2, we noted that the descriptions of the prevention and mitigation strategies suggested that they applied to different phases of the requirements lifecycle.
To capture this distinction, we further characterized the prevention and mitigation strategies based on the phases of the requirements lifecycle in which they occurred.
The first author assigned each low-level category from Step 2 to one or more of the requirements activities described by Kotonya and Sommerville~\cite{kotonya1998requirements} (see Table~\ref{tab:tblRequirementsActivities})
based on when they were most likely to occur.
The other authors reviewed this assignment, discussed any disagreements, and resolved them with full agreement.

\begin{table}[!htb]
    \caption{Requirements Activities in Software Development~\cite{hu2018using}}
    \label{tab:tblRequirementsActivities}
     \begin{tabular}{|p{.35\textwidth} | p{.65\textwidth} | }
  \hline
\textbf{Requirements Activities} & \textbf{Description} \\ 
  \hline
Requirements Elicitation & Activity that includes requirement gathering, understanding, organizing, and interacting with clients\\
  \hline
Requirements Analysis & Activity that includes requirements classification, prioritization, and resolving any misunderstanding\\
  \hline
Requirements Specification & Activity that includes defining user and system requirements and implementation, and ways to test and verify the implementation. It includes an activity of writing down the information gathered during the elicitation and analysis activities into a document that defines a set of requirements (System Specification Requirements Document). \\
  \hline
Requirements Validation & Activity that includes verification and validation of requirements specified in the specification activity. Ambiguity, consistency and redundancy check.
\\
  \hline
Requirements Management & Activity that includes management plans, resource collection and allocation, control to the changes in system requirements and so on.\\
  \hline
  
   \end{tabular}
\end{table}

\begin{figure}[!ht]
  \centering
  \includegraphics[width=5in]{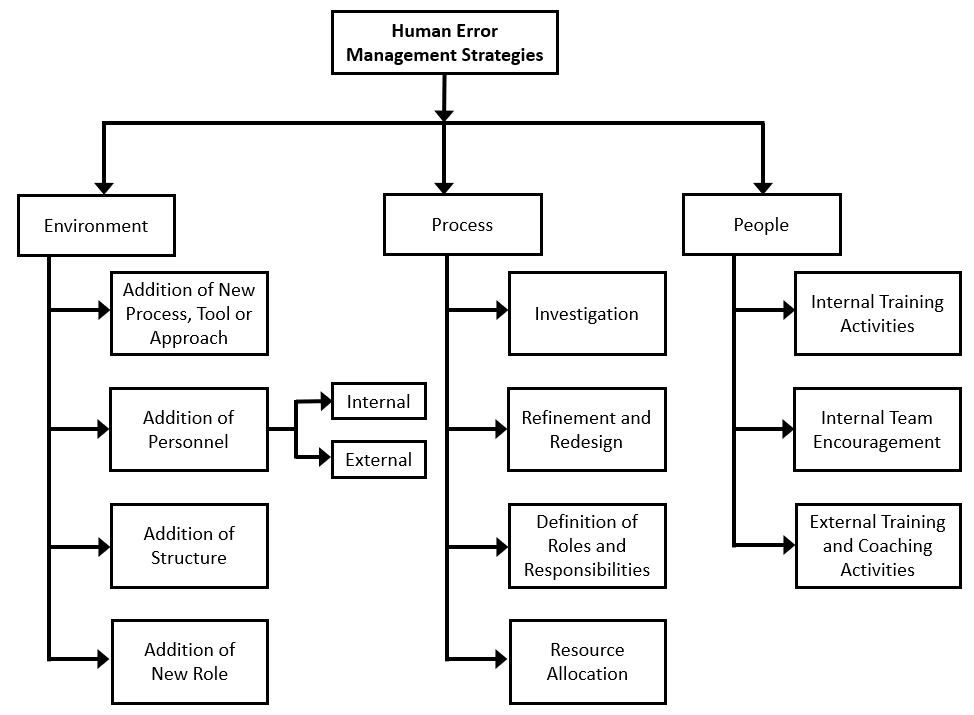}
  \caption{Human Error Management Taxonomy}
  \label{fig:PmTaxonomy}
\end{figure}

We organized the identified high-level classes and low-level sub-categories into the Human Error Management Taxonomy (HEMT) as shown in Figure~\ref{fig:PmTaxonomy}.

\subsection{RQ2: Can we categorize the human error prevention and mitigation strategies into the taxonomy developed in RQ1?}
\label{sec:RQ2}

To answer this question, we conducted \textbf{Phase 2: Classification of Prevention and Mitigation Strategies} of the analysis process. 
The goal of this phase was to classify each prevention and mitigation strategy into the high-level categories, the low-level categories, and the requirements activities defined during Phase 1.
Phase 2 has three Steps that build upon each other. 
We detail the steps and their results below.

\vspace{8pt}
\noindent
\textit{Step 1: Classify the prevention and mitigation strategies into the high-level classes}

We individually classified each prevention and mitigation strategy into the high-level classes defined in Step 1 of Phase 1.
To facilitate the process we randomly assigned the 162 prevention and mitigation strategies into five batches of 40, 31, 36, 40, and 15 strategies.
To the last batch, we added 13 randomly selected prevention and mitigation strategies that we had reviewed in previous batches. 
These strategies all had disagreements the first time so we wanted to validate the earlier classification of those strategies.
After individually classifying the strategies in each batch, we compared our results and discussed any disagreements to resolve them.
We identified 17 disagreements in the first batch, 13 in the second, 7 in the third, 8 in the fourth, and 8 in the fifth.
After discussing the disagreements, all three authors agreed on the final classification.
At this point, we merged two strategies.
We also classified eight strategies as Ambiguous and did not consider them for further classification.
These changes left 153 strategies for the next step.

Table~\ref{tab:tblPmRaType} shows the distribution of the prevention and mitigation strategies among the high-level classes and low-level sub-categories for each requirements activities to answer the research question. 
Following the methodology described above, we began by classifying the prevention and mitigation strategies into the  \textit{People}, \textit{Process}, and \textit{Environment} classes defined in response to RQ1.
During this process, we observed that most (51\%) of the prevention and mitigation strategies were \textit{Process} changes.
\textit{People} changes and \textit{Environment} changes made up 21\% and 23\% of the total respectively.
Eight strategies did not provide sufficient information, so we classified them as \textit{Ambiguous}.
The column \textbf{Total} in Table~\ref{tab:tblPmRaType} shows the total number of strategies that lie under each of the classes.

%Comment R4.12
\begin{table*}[!htb]
\centering
    \caption{Prevention and Mitigation Strategies Among Low-level Categories and Requirements Activities}
    \label{tab:tblPmRaType}
    \scriptsize
    \rotatebox{90}{
\begin{tabular}{|l|p{.2\textwidth}|l|l|l|l|l|l|l|}
\hline
\multicolumn{1}{|c|}{\multirow{2}{*}{\textbf{\makecell{P/M \\Classes}}}} & \multicolumn{1}{c|}{\multirow{2}{*}{\textbf{\makecell{P/M \\Sub-Categories}}}} & 
\multicolumn{6}{c|}{\textbf{Requirement Activities}} & \multirow{2}{*}{\textbf{Total}} \\ \cline{3-8}
\multicolumn{1}{|c|}{} & \multicolumn{1}{c|}{} & \textbf{Elicitation} & \textbf{Analysis} & \textbf{Specification} & \textbf{Validation} & \textbf{Management} & \textbf{\makecell{Unknown/ \\out of \\scope}} & 
\\ \hline
\multirow{6}{*}{\textbf{Environment}} & \makecell{Addition of New \\Process, Tool \\or Approach} & 0    & 6   & 0     & 5    & 7  & 0  & 18   \\ \cline{2-9} 
 & \makecell{Addition of \\Personnel\\ (Internal)} & 0  & 2 & 0  & 0 & 0 & 0 &2 \\ \cline{2-9}
 & \makecell{Addition of \\Personnel \\(External)} & 2 & 0 & 0 & 0 & 4 & 0 & 6 \\ \cline{2-9} 
& \makecell{Addition of \\ Structure} & 1 & 1 & 4 & 2 & 0 & 1 & 9 \\ \cline{2-9} 
& \makecell{Addition of \\ New Role} & 0 & 0 & 0 & 0 & 1 & 0 & 1 \\ \cline{2-9}
& \makecell{\textbf{Sub-total}} & \textbf{3} & \textbf{9} & \textbf{4} & \textbf{7} & \textbf{12} & \textbf{1} & \textbf{36}\\ \hline
\multirow{5}{*}{\textbf{Process}} & \makecell{Investigation} & 0 & 1 & 0 & 0 & 6 & 0 & 7 \\ \cline{2-9} & \makecell{Refinement \\ and Redesign} & 13 & 8 & 4 & 15 & 27 & 3 & 70 \\ \cline{2-9} 
 & \makecell{Definition of \\Roles and \\ Responsibilities} & 0 & 0 & 0 & 0 & 2 & 0 & 2 \\ \cline{2-9}
& \makecell{Resource \\Allocation}  & 0 & 0 & 0 & 0 & 0 & 0 & 2 \\ \cline{2-9}
& \makecell{\textbf{Sub-total}} & \textbf{15} & \textbf{9} & \textbf{4} & \textbf{15} & \textbf{35} & \textbf{3} & \textbf{81} \\ \hline
\multirow{4}{*}{\textbf{People}}  & \makecell{Internal \\ Training \\Activities} & 0 & 0 & 0 & 0 & 26 & 0 & 26 \\ \cline{2-9} 
& \makecell{Internal Team \\ Encouragement}  & 0 & 2 & 0 & 0 & 2 & 0 & 4 \\ \cline{2-9} 
 & \makecell{External Training \\and Coaching \\Activities} & 2 & 0 & 0 & 0 & 2 & 0 & 4\\ \cline{2-9}
& \makecell{\textbf{Sub-total}} & \textbf{2} & \textbf{2} & \textbf{0} & \textbf{0} & \textbf{30} & \textbf{0} & \textbf{34}  \\ \hline
\multicolumn{2}{|l|}{\textbf{Total}} & \textbf{20} & \textbf{20}  & \textbf{8} & \textbf{22} & \textbf{77} & \textbf{4} & \textbf{151} \\ \hline
\end{tabular}
}
\end{table*}

\vspace{8pt}
\noindent
\textit{Step 2: Classify the prevention and mitigation strategies into the low-level classes and requirements phases}

We further classified the strategies into low-level categories for each class.
The low-level categories denote the specific \textit{People}, \textit{Process}, and \textit{Environment} changes necessary to address the errors in requirements problems. 
The largest category, with 70 strategies is \textit{Process: Refinement and Redesign (RR)}.
One reason why this low-level category is so large is because it includes strategies that suggest changes in communication, documentation, review, and other processes as explained in Section~\ref{sec:Discussion}.
The next largest category is \textit{People: Internal Training Activities (ITA)} with 26 strategies, followed by \textit{Environment: Addition of New Process, Tool, or Approach (ANP)} with 18 strategies.
Although, ITA has more strategies than ANP, we found that overall there are not many low-level categories focused on people.
Moreover, most of the People strategies are related to training, likely because to implement changes in \textit{Process} or \textit{Environment}, people need to be trained to implement those changes.
Therefore, the results suggest it is more important to focus on changes in the \textit{Process} and the \textit{Environment} than the changes in \textit{People} to effectively prevent and mitigate human errors in requirements engineering.

Next, to help requirements engineers locate the activity where they should implement a change, we allocated the prevention and mitigation strategies across the requirements engineering lifecycle. 
Using the results from Step 1 as a starting point, the first author classified the prevention and mitigation strategies into the low-level classes.
She also classified each strategy into the requirements engineering lifecycle phase that was impacted by the change.
Using existing sources~\cite{kotonya1998requirements,parviainen2003requirements}, we defined the phases of the requirements engineering lifecycle as:

\begin{enumerate}
\item \textit{Elicitation}: 
includes requirements gathering, understanding, organizing, and interacting with clients;
\item \textit{Analysis}: 
includes requirements classification, prioritization, and resolving any misunderstanding;
\item \textit{Specification}: 
includes defining and documenting user and system requirements, the steps to implement the requirements, and ways to test and verify the implementation; 
\item \textit{Validation}: 
includes verification and validation of specified requirements to check for ambiguity, consistency, and redundancy;
\item \textit{Management}: 
includes the overall management activities such as management plans, resource collection and allocation, and change management; and
\item \textit{Unknown}: 
includes strategies that do not clearly state the relevant requirements activity.
\end{enumerate}

Then, the other two authors reviewed the classification and noted any disagreements.
Similar to the way we performed Step 1, we divided the 153 remaining strategies into 4 batches of 15, 30, 49, and 59 strategies respectively.
We had a total of 29 disagreements across all four batches, which we resolved through the discussion sessions.
At this point, we removed two strategies that still lacked sufficient information for analysis.
This change left 151 strategies for us to map onto the requirements activities.

Of these 151 strategies, we mapped 51\% to the \textit{Management} activity, 15\% to the \textit{Validation} activity, 13\% each to the \textit{Elicitation} and the \textit{Analysis} activities, and 5\% to the \textit{Specification} activity.
Only 3\% (or 4) of the strategies fell into the \textit{Unknown} category.

To summarize the results, we provide a complete list of the reported strategies in an online appendix\footnote{ http://carver.cs.ua.edu/Data/HumanError/Prevention-Mitigation-Strategies.html}.
For each strategy, the appendix lists the associated requirements problem, the underlying error type it addresses, and whether it is a prevention strategy or a mitigation strategy. 

\section{Discussion}
\label{sec:Discussion}
This section highlights key observations from the results of the study.

\subsection{Classifying the Strategies}
While analyzing the strategies, we noticed that due to the lack of detail provided and the subtle differences among some of them, it was not always easy to classify the strategies.
For example, a change in \textit{Environment} typically also requires a change in \textit{Process}, and to understand a change in \textit{Process} often requires a change in \textit{People}. 
However, we were able to classify most of the strategies to individual classes, placing only a small number in the \textit{Ambiguous} class.

It is interesting to note that all of the strategies analyzed in Section~\ref{sec:Results} required the \textit{introduction or definition} of processes, roles, or responsibilities.
Based on previous work that found the \textit{removal} of sub-tasks to be a strategy to address errors~\cite{huang2016post}, we expected to also see strategies that required the removal of process steps, roles, or responsibilities.
However, none of the strategies in our dataset focused on the removal of these items.
While it is possible that the `refinement' activities modified processes by removing steps or that `introduction' activities also implicitly included a removal of an existing activity, the dataset did not have any explicit mention of removal steps.
This point is something that is worth further investigation.

\subsection{Research Implications}
Table~\ref{tab:tblPmRaType} shows the number of strategies reported for each of the requirements activities.  
Out of a total of 159 strategies, 8 belonged to the Ambiguous class (not shown in the table), and 151 belonged to the People, Process, and Environment classes.
class), 
Of the 151 strategies, 81 belonged to the \textit{Process} class. 
In other words, slightly more than half of the time practitioners handle requirements errors by implementing changes to \textit{Process}. 
Among the strategies in the \textit{Process} class, ``Planning of regular communication events/meetings'' and ``Execution of regular communication events/meetings'' were the most common error management mechanisms, suggesting that a large number of requirements problems can be addressed through better communication.
The next most common strategy in the \textit{Process} class was ``Use of formal reviews,'' suggesting that requirements validation and review is an important practice for avoiding requirements engineering errors. 
Moreover, the fact that the second most common requirements engineering activity in which error management occurs was \textit{validation}, supports this implication. 

By distributing the strategies across the requirements engineering activities, our results help practitioners better identify which error management strategies are most appropriate for each phase.
As Table~\ref{tab:tblPmRaType} shows, 77 strategies (the largest set) fit into the \textit{Management} activity and 22 strategies into the \textit{Validation} activity.
In the case of the \textit{Management} activity, these strategies are primarily changes in \textit{Process} and changes in \textit{People}.
However, in the case of the \textit{Validation} activity, there are no strategies related to changes in \textit{People}.
Table~\ref{tab:tblPmRaType} shows that only 30 out of 77 strategies applied during the \textit{Management} activity focused on changes in \textit{People}.
Furthermore, there are either no or very few people-related strategies applied during other requirement activities.
The people-related strategies skew towards the management activity, which only happened because many of the management activities, such as organizing training and meetings, are related to changes in \textit{People}.
These observations show that requirements engineers manage human errors mostly with strategies related to changes in \textit{Process} and \textit{Environment} across the requirement activities.
Relating this observation back to the literature from the medical domain described in Section~\ref{sec:Background}, our results support the claim that changing people is not an effective way to manage human error~\cite{kellogg2017our}.  

The next three most common requirements engineering activities in which error management occurs are \textit{elicitation}, \textit{analysis} and, \textit{specification}.
Most of the strategies applied during elicitation are in the \textit{Process: Refinement and Redesign} low-level category.
The strategies for the analysis and specification activities equally focus on changes in \textit{Process} and \textit{Environment}.
This distribution of strategies again suggests that practitioners focused on changes in \textit{Process} and \textit{Environment} rather than change in \textit{People}.

Table~\ref{tab:tblPmRaType} shows the most common type of change in error management is \textit{Process: Refinement and Redesign (RR)}.
Because this category has a large number of strategies in it, we attempted to divide it further.
However, due to limited and vague nature of the data, we were not able to decompose it further. 
The fact that professional requirements engineers mostly apply strategies in RR class suggests that they believe these strategies are effective and successful.

Apart from RR, the next most common set of strategies is \textit{People: Internal Training Activities (ITA)}.
This result suggests that practitioners believe they can best manage most of the errors related to \textit{People} by improving skills through training. 
The other two low-level categories in the \textit{People} class contain only a small number of prevention and mitigation strategies. 

The next most common type of change is \textit{Environment: Addition of New Process, Tool or Approach (ANP)}.
Providing structure for tasks is another good error management mechanism in the \textit{Environment} class. 
The key observation here is that, while many requirements errors can be resolved by fixing the existing processes (as described above), sometimes there is a need for new processes.
The introduction of a new process often will require changes in existing processes which requires training the people to adapt the change. 

We also observed that the prevention and mitigation strategies in the \textit{Environment} class are somewhat evenly distributed among its low-level categories, as shown in Table~\ref{tab:tblPmRaType}. 
However, in the \textit{People} class, the strategies appear in only a small number of the low-level categories, with most appearing in \textit{People: Internal Training Activities (ITA)}. 
This distribution suggests that, to manage human errors, practitioners should focus on other important types of changes, like changes in \textit{Environment} and \textit{Process}, as opposed to changes in \textit{People}. 
Based on the literature from other domains (Section~\ref{sec:Background}), this implication is a positive sign~\cite{gosbee2003human,kellogg2017our,wu2008effectiveness}. 
A study from the medical domain that investigated how to manage human errors found solutions focused on changing people were seldom effective in handling the errors~\cite{kellogg2017our}.
This study also indicated that the error management approaches focused on changes in technologies, tools, and policies are more useful, yet less common~\cite{kellogg2017our}. 
Our analysis and observations derive similar implications.
Hence, our results answer the question in our title \textit{Should we fix the People, the Processes, or the Environment?} by suggesting to give priority to fixing the processes.

\subsection{Observed Features and Dimensions}
We observed several features exhibited by the RR subcategory in the \textit{Process}, including communication, documentation, planning and organization, and review and cross checks.
This observation means that most of the time, requirements engineers can overcome requirements errors and problems by redesigning or redefining the communication, documentation, planning, or reviewing processes.
During the analysis of the strategies, we attempted to divide the large RR category into smaller categories.
However, we found that we could identify more detailed distinct categories among the strategies.
For example, ``Use of check lists for monitoring requirements along their life cycles'' can be understood to have a ``documentation'' feature for maintaining checklists as well as a ``review and crosscheck'' feature for monitoring requirements.
Therefore, we did not further decompose the RR category because the information in our dataset did not provide sufficient detail for further analysis. 

During our analysis, we also determined that we could view the prevention and mitigation strategies through the perspectives of \textit{who}, \textit{what}, and \textit{how}. 
The \textit{who} perspective focuses on who instigates the change and who executes the change.
The \textit{what} perspective focuses on the change itself.
The \textit{how} perspective focuses on how the change is implemented.
We anticipate these perspective will be helpful for practitioners to understand who should take the responsibility for implementing the necessary changes.
Because the data in our sample did not have enough detail to clearly identify these perspectives, we were not able to add this dimension to our work.
However, we do think it will be a valuable direction to explore in our future work.

\subsection{Comparison of RET, HET and HEMT}
We present an overview of the three taxonomies - \textit{Requirements Error Taxonomy (RET)}, \textit{Human Error Taxonomy (HET)}, and \textit{Human Error Management Taxonomy (HEMT)}- in this sub-section.
RET provided a starting point to focus on errors as the root cause of faults in software~\cite{walia2009systematic}.
HET enhances the research in RET to investigates and categorize the errors from human cognitive perspective~\cite{anu2018development}.
In the current study, HEMT follows on the research in HET to introduce error management taxonomy and support error prevention and mitigation for the error types in HET. 
Table~\ref{tab:tblTaxonomyComparision} shows the comparison between the aforementioned taxonomies.

\begin{table}
 \caption{Comparison of RET, HET and HEMT}
 \label{tab:tblTaxonomyComparision}
 \footnotesize
 \rotatebox{90}{
\begin{tabular}{| p{0.15\textwidth} || p{0.35\textwidth} | p{0.35\textwidth} | p{0.35\textwidth}|}
\hline  
\textbf{Aspect} & \textbf{RET} & \textbf{HET}& \textbf{HEMT}\\
  \hline
Focus & Types of requirements errors that affect overall software quality. & Human errors based on classification system from Cognitive Psychology. & Human error prevention and mitigation approaches in requirements engineering based on HET.\\
  \hline
Approach & Bottom-up approach without a guiding error theory. & Extension of the RET People class with detailed categorization of human errors in requirements engineering & Complements HET by providing formal taxonomy of human error management classes and sub-categories\\
  \hline
Organization & Organizes requirements errors into People, Process, and Documentation categroies. & Organizes human errors in requirements engineering into Slips, Lapses, and Mistakes~\cite{reason1990human}. & Organizes human error management strategies into changes to People, Processes, and Environment.\\
  \hline
Construction Process & Systematic literature review of software engineering research. & Systematic Literature review of software engineering and cognitive psychology research. & Analysis of survey data from practitioners reporting actual experiences by the practitioners.\\
  \hline
Use & Supports fault detection, error detection, and error prevention for requirements errors. Because these errors are not true human errors, errors can still occur resulting in faults. & Supports fault detection, error detection, and error prevention. The human error information in the HET will help requirement engineers avoid making such errors and hence prevent the faults caused by such errors. & Supports error prevention and error mitigation. It provides requirements engineers with knowledge about strategies to prevent and mitigate different types of errors.\\

\hline  
\end{tabular}
}
\end{table}

\section{Threats to Validity}
\label{sec:Threats}
This section describes the threats to validity for the study and how we addressed them.

\subsection{Internal Validity} 
The survey data is subjective. 
We do not know whether the reported strategies were useful nor how confident the respondents were in their answers. 
However, because the participants were experienced industrial developers from different software industries and different countries, we believe this threat is minimal relative to our goals in this paper. 
Moreover, we discarded any responses that contained either vague explanations or explanations unrelated to our research questions.
To prevent potential confirmation bias from including only those responses that confirmed our goal, all three authors separately analyzed the data and agreed upon the items we eliminated.
However, we do need further studies to evaluate the usefulness of the reported strategies.

\subsection{Construct Validity}
For the NaPiRE dataset, the respondents provided information about how they prevent and mitigate requirements problems.
However, the survey was not specifically focused on human errors.
Therefore, in earlier work, two of the authors mapped these problems to human errors~\cite{hu2018using}.
It is possible that either the previous work or our current work misinterpreted the information provided on the survey because we did not interact directly with the respondents.
Furthermore, some of the responses were not in English and others were ambiguous.
To minimize the threat, we only consider responses in English and created a separate class for the ambiguous responses.

For the CAPS survey, the participants first received training about human errors in requirements engineering before answering survey questions about prevention and mitigation strategies. 
Even so, it is possible they did not fully understand human errors and therefore provided irrelevant information. 

To help address these threats, we analyzed the reported strategies along with other information on the survey about the problem causes and the error type to help us properly interpret the responses. 
The three authors each performed the analysis separately and then met to resolve any disagreements. 
These different perspectives also reduces the potential for this threat.

\subsection{External Validity} 
Due to its design, the information provided by the NaPiRE survey provides a broad cross-section of software organizations from around the world.
This diversity increases the likelihood that the reported results have external validity.
Conversely, the data from the CAPS survey represents information from one small organization.
We used data from both surveys to draw conclusions in the paper.
Because the main focus of the study was to propose a taxonomy based on data from industrial practitioners, the threat to external validity is low.
Furthermore, the taxonomy is built based on actual error management strategies used by practitioners to manage real requirements problems.
As a result, the threat to external validity concerning whether the taxonomy captures real strategies is low.
However, we do note that building the taxonomy is the first step in a larger research endeavor which now requires additional studies to validate (and improve) the taxonomy.

\section{Conclusion}
\label{sec:Conclusion}
It is difficult to identify human errors that occur during the requirements engineering process because they often manifest themselves much later as faults and may remain hidden until execution. 
Practitioners need to address human errors as early as possible in the software development process to reduce cost and improve software quality. 
The lack of research on human error in requirements engineering makes it difficult for requirements engineers to understand the existence of these human errors and the techniques to manage them. 
This study on error management mechanisms derived from industrial practice provides an initial taxonomy of error management methods. 

The results of our analysis produced three high-level classes of prevention and mitigation strategies: \textit{People}, \textit{Process}, and \textit{Environment}.
Inside each of these high-level classes, we also identified a set of lower-level classes to further distribute the strategies. 
Our resulting taxonomy provides some structure around human error management approaches in requirements engineering.
Future research can use this structure as a starting point to evaluate the effectiveness of each prevention and mitigation strategy and to identify classes missing from the taxonomy.
In addition, the distribution of error management approaches across the requirements activities will help researchers and practitioners gain a better understanding of which approaches are most applicable during which activities.

The study also provided some interesting insights into the error management approaches used in industry. 
Though human error is defined as error in human cognition and is generally related to people, most of the error management strategies identified in the surveys are focused on changes in \textit{process} rather than changes in \textit{people}.
Based on the prevalence of strategies in the the \textit{process} class and the fact that those strategies cover most types of human errors, we can infer that process-related error management strategies may be more useful than those from the other classes.
This observation is consistent with the research from the medical domain that found that people-related solutions are less successful than process-related solutions~\cite{kellogg2017our}.

This paper describes an initial study on the topic of human error management in requirements engineering, specifically, and software engineering, more generally. 
We built the taxonomy from requirements problems and solutions described by practitioners.
Therefore, the taxonomy itself has value as a systematization of this information and as a base for additional research exploring human error management in software engineering.
All of the items in the taxonomy, including each low-level category, need to be investigated with studies designed with more specific goals related to the prevention and mitigation strategies.

In our future work, we plan to evaluate the effectiveness, usefulness, and completeness of the taxonomy in an industrial setting.
To do this evaluation, we intend to have practitioners use the taxonomy to help them prevent or mitigate the problems they face in their projects.
Then, we will investigate how well the error management strategies in the taxonomy help practitioners overcome the requirements problems they face.
We expect that, through empirical studies, we can compare the effectiveness of the types of prevention and mitigation strategies to identify the most effective ones to guide practitioners.

\bibliographystyle{elsarticle-num} 
\bibliography{ref}

\begin{thebibliography}{10}
\expandafter\ifx\csname url\endcsname\relax
  \def\url#1{\texttt{#1}}\fi
\expandafter\ifx\csname urlprefix\endcsname\relax\def\urlprefix{URL }\fi
\expandafter\ifx\csname href\endcsname\relax
  \def\href#1#2{#2} \def\path#1{#1}\fi

\bibitem{reason1990human}
J.~Reason, Human error, Cambridge university press, 1990.

\bibitem{viller1997human}
S.~Viller, J.~Bowers, T.~Rodden, Human factors in requirements engineering,
  Tech. rep., Technical Report, Cooperative Systems Engineering Group,
  CSEG/8/1997 (1997).

\bibitem{askarinejadamiri2017impact}
Z.~Askarinejadamiri, H.~Zulzallil, A.~A.~A. Ghani, K.~T. Wei, Impact
  propagation of human errors on software requirements volatility, Int. J. Adv.
  Comput. Sci. Appl. 8~(2) (2017) 227--237.

\bibitem{hu2018using}
W.~Hu, J.~C. Carver, V.~Anu, G.~S. Walia, G.~L. Bradshaw, Using human error
  information for error prevention, Empirical Software Engineering 23~(6)
  (2018) 3768--3800.

\bibitem{fernandez2017naming}
D.~M. Fern{\'a}ndez, S.~Wagner, M.~Kalinowski, M.~Felderer, P.~Mafra,
  A.~Vetr{\`o}, T.~Conte, M.-T. Christiansson, D.~Greer, C.~Lassenius, et~al.,
  Naming the pain in requirements engineering, Empirical software engineering
  22~(5) (2017) 2298--2338.

\bibitem{kitchenham2010can}
B.~Kitchenham, D.~I. Sj{\o}berg, O.~P. Brereton, D.~Budgen, T.~Dyb{\aa},
  M.~H{\"o}st, D.~Pfahl, P.~Runeson, Can we evaluate the quality of software
  engineering experiments?, in: Proceedings of the 2010 ACM-IEEE International
  Symposium on Empirical Software Engineering and Measurement, 2010, pp. 1--8.

\bibitem{anu2018development}
V.~Anu, W.~Hu, J.~C. Carver, G.~S. Walia, G.~Bradshaw, Development of a human
  error taxonomy for software requirements: a systematic literature review,
  Information and Software Technology 103 (2018) 112--124.

\bibitem{gosbee2003human}
J.~Gosbee, T.~Anderson, Human factors engineering design demonstrations can
  enlighten your rca team, BMJ Quality \& Safety 12~(2) (2003) 119--121.

\bibitem{kellogg2017our}
K.~M. Kellogg, Z.~Hettinger, M.~Shah, R.~L. Wears, C.~R. Sellers, M.~Squires,
  R.~J. Fairbanks, Our current approach to root cause analysis: is it
  contributing to our failure to improve patient safety?, BMJ quality \& safety
  26~(5) (2017) 381--387.

\bibitem{wu2008effectiveness}
A.~W. Wu, A.~K. Lipshutz, P.~J. Pronovost, Effectiveness and efficiency of root
  cause analysis in medicine, Jama 299~(6) (2008) 685--687.

\bibitem{hendrick2002ergonomic}
H.~W. Hendrick, Ergonomic design of controls, displays, and workspace
  arrangements to reduce human error, in: Proceedings Kongres Nasional XI dan
  Seminar Ilmiah XIII Ikatan Ahli Ilmu Faal Indonesia dan International Seminar
  on Ergonomics and Sports Physiology, 2002, pp. 14--17.

\bibitem{huang2012taxonomy}
F.~Huang, B.~Liu, B.~Huang, A taxonomy system to identify human error causes
  for software defects, in: The 18th international conference on reliability
  and quality in design, 2012, pp. 44--49.

\bibitem{huang2017software}
F.~Huang, L.~Bin, Software defect prevention based on human error theories,
  Chinese Journal of Aeronautics 30~(3) (2017) 1054--1070.

\bibitem{huang2016post}
F.~Huang, Post-completion error in software development, in: 2016 IEEE/ACM
  Cooperative and Human Aspects of Software Engineering (CHASE), IEEE, 2016,
  pp. 108--113.

\bibitem{back2007does}
J.~Back, W.~L. Cheng, R.~Dann, P.~Curzon, A.~Blandford, Does being motivated to
  avoid procedural errors influence their systematicity?, in: People and
  Computers XX—Engage, Springer, 2007, pp. 151--157.

\bibitem{firesmith2007common}
D.~Firesmith, Common requirements problems, their negative consequences, and
  the industry best practices to help solve them., Journal of Object Technology
  6~(1) (2007) 17--33.

\bibitem{lopes2013application}
M.~E. R.~F. Lopes, C.~H.~Q. Forster, Application of human error theories for
  the process improvement of requirements engineering, Information Sciences 250
  (2013) 142--161.

\bibitem{walia2009systematic}
G.~S. Walia, J.~C. Carver, A systematic literature review to identify and
  classify software requirement errors, Information and Software Technology
  51~(7) (2009) 1087--1109.

\bibitem{lanubile1998experimenting}
F.~Lanubile, F.~Shull, V.~R. Basili, Experimenting with error abstraction in
  requirements documents, in: Proceedings Fifth International Software Metrics
  Symposium. Metrics (Cat. No. 98TB100262), IEEE, 1998, pp. 114--121.

\bibitem{walia2010evaluating}
G.~S. Walia, J.~C. Carver, Evaluating the use of requirement error abstraction
  and classification method for preventing errors during artifact creation: A
  feasibility study, in: 2010 IEEE 21st International Symposium on Software
  Reliability Engineering, IEEE, 2010, pp. 81--90.

\bibitem{kotonya1998requirements}
G.~Kotonya, I.~Sommerville, Requirements engineering: processes and techniques,
  Wiley Publishing, 1998.

\bibitem{parviainen2003requirements}
P.~Parviainen, H.~Hulkko, J.~K{\"a}{\"a}ri{\"a}inen, J.~Takalo, M.~Tihinen,
  Requirements engineering: Inventory of technologies, VTT Technical Research
  Centre of Finland, 2003.

\end{thebibliography}

\appendix
\clearpage
\section{Requirements Problems Reported in NaPiRE Dataset}
\label{appendixA}
\begin{table*}[!htb]
\centering
    %\caption{21 Requirements Problems}
    \label{tab:tblReqProb}
    %\scriptsize
% \usepackage{longtable}
% Note: It may be necessary to compile the document several times to get a multi-page table to line up properly
 \begin{tabular}{|p{.2\textwidth} | p{.8\textwidth} | }
  \hline
\textbf{Problems No.} & \textbf{Comment Requirements Problems} \\ 
  \hline
1              & Communication flaws within the project team                                    \\ \hline
2              & Communication flaws between us and the customer                                \\ \hline
3              & Terminological problems                                                        \\ \hline
4              & Unclear responsibilities                                                       \\ \hline
5              & Incomplete and / or hidden requirements                                        \\ \hline
6              & Insufficient support by project lead                                           \\ \hline
7              & Insufficient support by customer                                               \\ \hline
8 & Stakeholders with difficulties in separating requirements from previously known solution designs \\ \hline
9              & Inconsistent requirements                                                      \\ \hline
10             & Missing traceability                                                           \\ \hline
11             & Moving targets (changing goals, business processes and / or requirements)      \\ \hline
12             & "Gold plating" (implementation of features without corresponding requirements) \\ \hline
13             & Weak access to customer needs and / or (internal) business information         \\ \hline
14             & Weak knowledge of customer's application domain                                \\ \hline
15             & Weak relationship to customer                                                  \\ \hline
16             & Time boxing / Not enough time in general                                       \\ \hline
17 & Discrepancy between high degree of innovation and need for formal acceptance of (potentially wrong / incomplete / unknown) requirements \\ \hline
18             & Technically unfeasible   requirements                                          \\ \hline
19 & Underspecified requirements that are too abstract and allow for various interpretations \\ \hline
20             & Unclear / unmeasurable non-functional requirements                             \\ \hline
21 & Volatile customer's business domain regarding, e.g., changing points of contact, business processes or   requirements\\ \hline
\end{tabular}
\end{table*}

\end{document}